\documentclass[12pt]{article}
\usepackage{graphicx}

\newcommand\pubnumber{WSU--HEP--XXYY}
\newcommand\pubdate{\today}

\def\ihep{Institute of High Energy Physics,Chinese Academy of Sciences, PO BOX 918(1), Beijing 100049}

\def  \jp {J/\psi}
\def \psip {\psi(3686)}
\def \psipp {\psi(3770)}
\def \ee {e^+e^-}
\def \etac {\eta_c}
\def \etacp {\eta_c(2S)}
\def \chicj {\chi_{cJ}}

\newcommand{\solutionA}{[2.94 \pm 0.16\textrm{(stat.)} \pm 0.16\textrm{(syst.)}] \times 10^{-6}}
\newcommand{\solutionB}{[1.24 \pm 0.33\textrm{(stat.)} \pm 0.30\textrm{(syst.)}] \times 10^{-7}}

\def\Title#1{\begin{center} {\Large #1 } \end{center}}
\def\Author#1{\begin{center}{ \sc #1} \end{center}}
\def\Address#1{\begin{center}{ \it #1} \end{center}}

\newcommand\pubblock{\rightline{\begin{tabular}{l} \pubnumber\\
         \pubdate  \end{tabular}}}
\newenvironment{Abstract}{\begin{quotation}  }{\end{quotation}}
\newenvironment{Presented}{\begin{quotation} \begin{center}
             PRESENTED AT\end{center}\bigskip
      \begin{center}\begin{large}}{\end{large}\end{center} \end{quotation}}
\def\Acknowledgements{\bigskip  \bigskip \begin{center} \begin{large}
             \bf ACKNOWLEDGEMENTS \end{large}\end{center}}

\def\beq{\begin{equation}}
\def\eeq#1{\label{#1}\end{equation}}
\def\eeqn{\end{equation}}

\def\beqa{\begin{eqnarray}}
\def\eeqa#1{\label{#1}\end{eqnarray}}
\def\eeqan{\end{eqnarray}}

\let\bar=\overbar

\def\Dslash{\not{\hbox{\kern-4pt $D$}}}
\def\dslash{\not{\hbox{\kern-2pt $\del$}}}

\def\ee{e^+e^-}

\def\msb{{\bar{\ssstyle M \kern -1pt S}}}

\begin{document}
\begin{titlepage}
\pubblock

\vfill
\Title{Studies of Charmonium at BESIII}
\vfill
\Author{Rong-Gang Ping}
\Address{\ihep}
\vfill
\begin{Abstract}
Based on $\psip$ decays of 106 million, the 1.31 billion $\jp$ events and a data sample of $\psi(3770)$ with $2.9~fb^{-1}$ integrated luminosity, many analyses are performed. Exclusively baryonic decays of the $\psipp$, the radiative transition $\psipp\to\gamma\etacp$, the $\psipp$ transition to $\chicj$, isospin violation decay $\chi_{c0,2}\to\pi^0\etac$, the $C-$parity violation decays $\jp\to\gamma\gamma,\gamma\phi$ are searched for, but no significant signals are observed, and upper limits are set for these decays. The decays of $\psipp\to\gamma\chi_{c1}$ and $\jp\to\pi^0\phi$ signals are observed. These measurements provide more information on the charmonium structure, and the isospin and $C$-parity violation in the charmonium decays.
\end{Abstract}
\vfill
\begin{Presented}
The 7th International Workshop on Charm Physics (CHARM 2015)\\
Detroit, MI, 18-22 May, 2015
\end{Presented}
\vfill
\end{titlepage}
\def\thefootnote{\fnsymbol{footnote}}
\setcounter{footnote}{0}
%

\section{Introduction}
Results in this presentation are based on data samples accumulated with the BESIII detector at the BEPCII collider, which include $\psip$ decays of 106 million, a sample of 1.31 billion $\jp$ events and $\psi(3770)$ data of $2.9~fb^{-1}$ integrated luminosity, and 42 $pb^{-1}$ continuum data taken at 3.65 GeV.

\section{Search for $\psipp$ exclusive decays}

The nature of the excited $J^{PC}=1^{--}~c\bar c$  bound states
above the $D\bar D$ threshold is of interest but still not well
known. The $\psipp$ resonance, as the lightest charmonium
state lying above the open charm threshold, is
generally assigned to be a dominant $1^3D_1$ momentum
eigenstate with a small $2^3S_1$ admixture \cite{SDmixing}. It has been
thought almost entirely to decay to $D\bar D$ final states \cite{DDbardecays}. Unexpectedly, the BES Collaboration found a large inclusive
non-$D\bar D$ branching fraction, $(14.7\pm3.2)$\%, by utilizing
various methods \cite{besnonddbar}, neglecting interference effects,
and assuming that only one $\psipp$ resonance exists in
the center-of-mass energy between 3.70 and 3.87 GeV. A
later work by the CLEO Collaboration taking into account
the interference between the resonance decays and continuum
annihilation of $e^+e^-$ found a contradictory non-$D\bar D$
branching fraction, $(-3.3\pm1.4^{+6.6}_{-4.8})$. The BES results
suggest substantial non-$D\bar D$ decays, although the CLEO
result finds otherwise. Till now the observed non-$D\bar D$ exclusive decays sum up to less than 2\% of all decays, which motivates the search for other exclusive non-$D\bar D$ final states.

\subsection{Baryonic decays of $\psipp$}
By analyzing data samples of 2.9 fb$^{-1}$ collected at $\sqrt s=3.773$ GeV, the exclusive decays to final states, $\Lambda \bar\Lambda\pi^+\pi^-$, $\Lambda
\bar\Lambda\pi^0$, $\Lambda \bar\Lambda\eta$, $\Sigma^+ \bar\Sigma^-$,
$\Sigma^0 \bar\Sigma^0$, $\Xi^-\bar\Xi^+$ and $\Xi^0\bar\Xi^0$, are searched for \cite{baryonicdecay}. The QED backgrounds are estimated with the data samples taken at $\sqrt s$=3.542,~3.554, 3.561, 3.600 and 3.650 GeV, and the backgrounds from the initial state radiation (ISR), e.g. $\ee\to \gamma\psip,~\gamma\psipp$ are estimated with Monte-Carlo (MC) simulation. After subtraction of these backgrounds, no significant signals are observed. The upper limits at the 90\% confidence level (CL) are set as $4.4,0.7,1.9,1.0,0.4,1.5$ and 1.4 ($\times 10^{-4}$) for $\psipp\to\Lambda \bar\Lambda\pi^+\pi^-$, $\Lambda
\bar\Lambda\pi^0$, $\Lambda \bar\Lambda\eta$, $\Sigma^+ \bar\Sigma^-$,
$\Sigma^0 \bar\Sigma^0$, $\Xi^-\bar\Xi^+$ and $\Xi^0\bar\Xi^0$, respectively. 

These results provide
useful information for understanding the nature of $\psi(3770)$, but the
large non-$D\bar D$ component still remains a puzzle. A fine energy
scan over $\psi(3770)$ and $\psi(4040)$ resonances would be very
helpful for obtaining the lineshape of exclusive non-$D\bar D$
processes, and help determine whether the processes exist or not.
\subsection{$\psipp\to\gamma\eta_c(2S)\to\gamma K_S^0 K\pi$}
The radiative transitions $\psipp\to\gamma\etacp$ are supposed to be highly suppressed by selection rules, considering the $\psipp$ is predominantly the $1^3D_1$ state. However, due to the non-vanishing photon energy in the decay, higher multipoles beyond the leading one could contribute~\cite{IML}. Experimental measurements of the branching fractions $\mathcal{B}(\psipp\to\gamma\etacp$ will be very helpful  for testing theoretical predictions and providing further constraints on the immediate meson loop calculation (IML) contributions.

\begin{figure}[htb]
\centering
\includegraphics[height=3in]{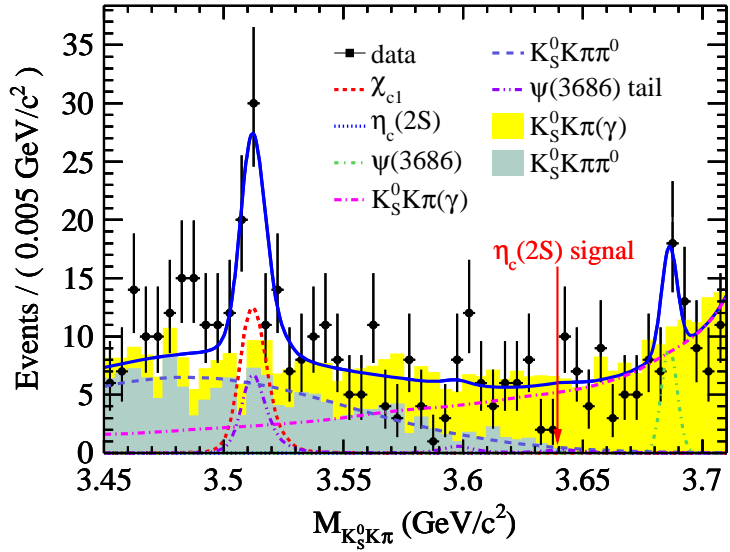}
\caption{Invariant-mass spectrum for $K^0_SK^\pm\pi^\mp$ from data with  the estimated backgrounds and best-fit results superimposed in the $\etacp$ mass regions.  Dots with error bars are data. The shaded histograms represent the background contributions. The solid lines show the total fit results.}
\label{psipp2getacp}
\end{figure}

Using the 2.92~fb$^{-1}$ data sample taken at $\sqrt{s} = 3.773$~GeV, searches for the radiative  transitions between the $\psipp$ and the $\etacp$ through the decay process  $\psipp\to\gamma K^0_SK^\pm\pi^\mp$ are performed \cite{psipp2getacp}. Figure \ref{psipp2getacp} shows the invariant mass distribution in the $\etacp$ mass region. No significant $\etacp$ signals are observed. Upper limits on the branching fraction at a 90\% C.L. is set as
$\psipp\!\to\!\gamma\etacp\to\gamma K^0_SK^\pm\pi^\mp)\!<\!5.6\!\times\!10^{-6},\textrm{~or~}
\mathcal{B}(\psipp\!\to\!\gamma\etacp)\!<\!2.0\!\times\!10^{-3}$.

The measured branching fraction corresponds to the partial decay width 55 KeV, which is larger than the prediction of the IML \cite{IML} ($0.6\sim$3.8 KeV). This is due to the limitation by statistics and the dominant systematic error, which stems from the uncertainty in the branching fraction of $\etacp\to K^0_SK^\pm\pi^\mp$.
\subsection{$\psipp\to\gamma\chicj$}
Within an $S$-$D$ mixing model, the $\psi(3770)$ resonance
is assumed to be predominantly the $1^3D_1$ $c\bar c$ state with a small admixture of the $2^3S_1$ state.
Based on this assumption, 
the partial widths of $\psi(3770)$ $E1$ radiative transitions are predicted with large uncertainties ~\cite{e1transition}. Precision measurements of partial widths of the
$\psi(3770)\to\gamma\chi_{c1,2}$ processes
are critical to test the above mentioned models, and
to better understand the nature of the $\psi(3770)$, as well as to find
the origin of the non-$D\bar D$ decays of the $\psi(3770)$.

\begin{figure}[htb]
\centering
\includegraphics[height=3in]{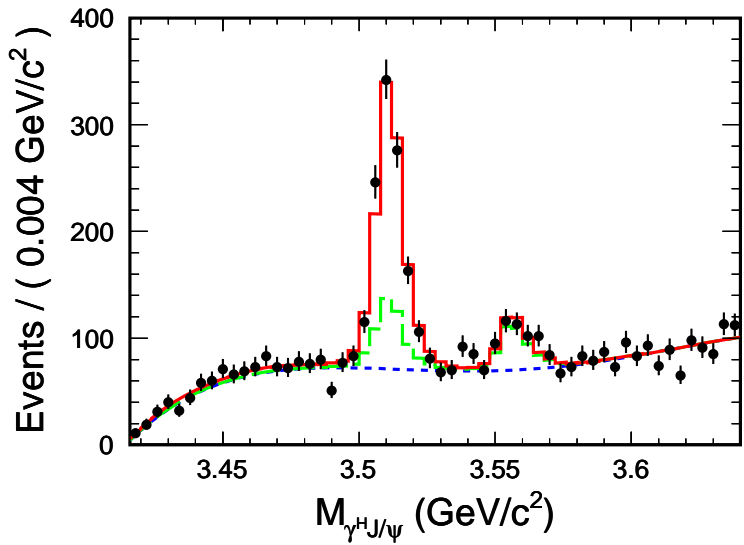}
\caption{Invariant mass spectrum of the energetic photon and $J/\psi$ combinations selected from data.
The dots with error bars represent the data.
The solid (red) line shows the fit.
The dashed (blue) line shows the smooth background.
The long-dashed (green) line is the sum of the smooth background
and the contribution from $e^+e^-\to(\gamma_{\rm ISR})\psi(3686)$ production.}
\label{psipp2gchicj}
\end{figure}

By analyzing 2.92~fb$^{-1}$ of data collected at $\sqrt{s}=3.773$~GeV, the decay $\psi(3770) \to \gamma \chi_{c1}$ was searched for \cite{psipp2gchicj}, and the $\chicj$ candidates are reconstructed with the decay $\chicj\to\gamma\jp$. Figure \ref{psipp2gchicj} shows the invariant mass spectrum of the energetic photon and $J/\psi$. The decays to $\chi_{c1}$ are observed, while no significant signal for the $\chi_{c2}$.  The branching fraction is measured to be
$\mathcal B(\psi(3770) \to \gamma \chi_{c1})=(2.48 \pm 0.15 \pm 0.23) \times 10^{-3}$
and a $90\%$ C.L. upper limit $\mathcal B(\psi(3770) \to \gamma \chi_{c2}) < 0.64 \times 10^{-3}$.
This measured branching fraction for $\psi(3770) \to \gamma \chi_{c1}$
is consistent within error with
$\mathcal B(\psi(3770) \to \gamma \chi_{c1})=(2.8 \pm 0.5\pm 0.4) \times 10^{-3}$
measured by CLEO-c~\cite{prl96_182002},
but the precision of this measurement is improved by more than a factor of 2.

\section{Searches for isospin-violating transitions $\chi_{c0,2}\to\pi^0\etac$}
Isospin is known to be a good symmetry in the
hadronic decays of charmonium states. The decay
rates of isospin-symmetry breaking modes are in general found to be very small. However, the isospin transition $\psip\to\pi^0\jp$ is observed with large branching fraction; the ratio $\mathcal{B}(\psip\to\pi^0\jp)/\mathcal{B}(\psip\to\eta\jp)$ is measured to be $0.374\pm0.072$ \cite{liaogr}, which indicates the important role played by the nonperturbative effects \cite{zhaoq}. Searches for the isospin decay $\chicj\to\pi^0\etac$ gives insights in the isospin-violating mechanisms.

An analysis was performed with the aim to search for
the hadronic isospin-violating transitions $\chi_{c0,2} \rightarrow \pi^{0} \eta_{c}$ using $106\times 10^{6}$
$\psi(3686)$ events collected by BESIII through $\eta_{c} \rightarrow K^{0}_{S}K^{\pm}\pi^{\mp}$ decays \cite{olga}.
No statistically significant signal is observed and upper limits on the branching fractions for the
processes $\chi_{c0,2} \rightarrow \pi^{0} \eta_{c}$ have been obtained. The results are
$B(\chi_{c0}\rightarrow \pi^0 \eta_c) < 1.6\times 10^{-3}$ and $B(\chi_{c2}\rightarrow \pi^0 \eta_c) < 3.2\times 10^{-3}$.
These are the first upper limits that have been reported so far.
These limits might help to constrain nonrelativistic field theories and provide insight in the role of charmed-meson loops
to the various transitions in charmonium and charmonium-like states.
Further developments in these theories will be necessary to clarify this aspect.

The obtained upper limit on $B(\chi_{c0} \rightarrow \pi^{0} \eta_{c})$ does not contradict the theoretical estimate of order (few)$\times$10$^{-4}$ ~\cite{Voloshin2012} .
In addition, the branching fractions of the
hadronic decays $\chi_{c0}\rightarrow \pi^0 \eta_c$ and $\chi_{c1}\rightarrow \pi^+\pi^-\eta_c$
are predicted approximately equal \cite{Voloshin2012}. An earlier theoretical estimate in the framework of a QCD multipole expansion~\cite{Lu2007}
reported a branching fraction for $\chi_{c1}\rightarrow\pi\pi\eta_c$ of (2.22$\pm$1.24)\%, which
contradicts the earlier BESIII measurement~\cite{BESIIIpipietac} and, under the above relation~\cite{Voloshin2012}, this measurement as well.
\section{Search for $C-$violation decay $\jp\to\gamma\gamma,\gamma\phi$}
In the Standard Model~(SM), $C$-invariance
is held in strong and electromagnetic~(EM) interactions.
Until now, no $C$-violating processes have been observed in EM
interactions~\cite{pdg14}. While both $C$-parity and $P$-parity can be
violated in the weak sector of the electroweak interactions in the SM,
evidence for $C$ violation in the EM sector would immediately indicate
physics beyond the SM.

\begin{figure}[htb]
\centering
\includegraphics[height=2.in]{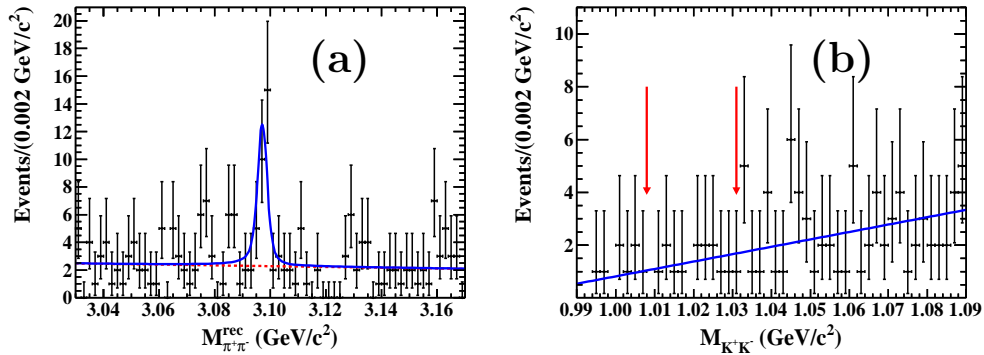}
\caption{(a) The $ M^{\rm rec}_{\pi^+\pi^-}$ distribution for
$\psi(3686) \to J/\psi \pi^+\pi^-, J/\psi \to \gamma\gamma$
candidate events from data. (b) The $ M_{K^+K^-}$ distribution for $\psi(3686) \to J/\psi \pi^+\pi^-, J/\psi \to \gamma\phi, \phi \to K^+K^-$
candidate events from data. The solid line shows the
global fit results and the dashed line shows the background.}
\label{jp2gg}
\end{figure}

Using $1.06\times10^8$ $\psi(3686)$ events recorded in $e^{+}e^{-}$ collisions at $\sqrt{s}=$ 3.686 GeV, we performed searches for the $C$-violation decays $J/\psi \to \gamma\gamma$ and $J/\psi \to \gamma\phi$ using transition $\psip\to\pi^+\pi^-\jp$ \cite{j2gg}. The $\phi$ candidates are reconstructed with the decay $\phi\to K^+K^-$. The signals for $\jp\to\gamma\gamma$ are searched for by looking for the $\jp$ candidates in the recoil mass distribution $M_{\pi^+\pi^-}$, while signals for $\jp\to\gamma\phi$ are searched for by looking for the $\phi$ candidates in the invariant mass $M_{K^+K^-}$ .
No significant signals are observed as shown in Fig. \ref{jp2gg}. We set the upper limits $\mathcal{B}(J/\psi \to \gamma\gamma)
< 2.7 \times 10^{-7}$ and $\mathcal{B}(J/\psi \to \gamma\phi)  < 1.4 \times 10^{-6}$ at the 90\% C.L. for
the branching fractions of $J/\psi$ decays into $\gamma\gamma$ and $\gamma\phi$, respectively.
The upper limit on
$\mathcal{B}(J/\psi \to \gamma\gamma)$ is one order of magnitude more
stringent than the previous upper limit, and $\mathcal{B}(J/\psi \to \gamma\phi)$ is the first
upper limit for this channel. Our results are consistent with $C$-parity conservation of the EM interaction.
\section{Search for the OZI-suppressed decay $\jp\to\pi^0\phi$}
The $\jp$ hadronic decays proceed via the $c\bar c$ quarks annihilation into gluons, and then they materialize into light hadrons. Thus the $\jp$ is characterized by the narrow decay width, which is known as the Okubo-Zweig-Iizuka (OZI) suppressed decay \cite{ozi}. A full investigation of $\jp$ decaying to a vector meson ($V$) and a
pseudoscalar meson ($P$) can provide rich information about SU(3) flavor
symmetry and its breaking, probe the quark and gluon content of the
pseudoscalar mesons, and determine the electromagnetic
amplitudes~\cite{qedamp}. For the $\jp\to\pi^0\phi$, the partial decay width is even more suppressed due to no quark correlation in the final states, which is regarded as double OZI (DOZI) decay.
Well established phenomenological models~\cite{phemodel} have
indicated that the DOZI amplitude can have a large impact through
interference with the singly OZI suppressed amplitude. To search for the decay $\jp \to \phi \pi^0$ is helpful for us to understand the electromagnetic DOZI mechanisms of non-ideal $\omega-\phi$
mixing~\cite{phemodel,opmix}.

\begin{figure}[htb]
\centering
\includegraphics[height=4in]{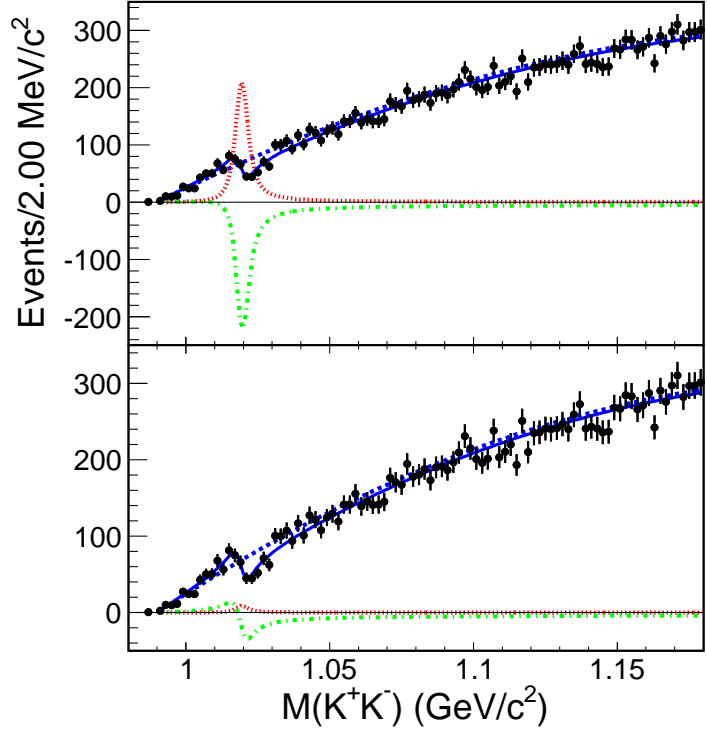}
\caption{Fit to $M(K^+K^-)$ spectrum after sideband subtraction for
  Solution I (a) and Solution II (b). The red dotted curve denotes the
  $\phi$ resonance; the blue dashed curve is the non-$\phi$
  contribution; the green dot-dashed curve represents their interference;
  and the blue solid curve is the sum of them.}
\label{j2piphi}
\end{figure}

Based on 1.31 billion $\jp$ events, we performed an analysis of the decay $\jp \to \phi
\pi^0 \to K^+K^- \gamma\gamma$ and find a structure around $1.02$~GeV/$c^2$
in the $K^+K^-$ invariant mass spectrum as shown in Fig. \ref{j2piphi}. It can be interpreted as
interference of $\jp \to \phi \pi^0$ with other processes decaying to
the same final state. The fit yields two possible solutions and thus
two branching fractions, $\solutionA$ and $\solutionB$ \cite{j2phipi}.

Using the measured branching fraction $\mathcal{B}(\jp \to
\phi \pi^0)$ and $\mathcal{B}(\jp \to
\omega \pi^0) = (4.5\pm0.5)\times10^{-4}$~\cite{pdg14}, one can extract the $\omega-\phi$ mixing parameters~\cite{phemodel}. If $\omega-\phi$
are mixed ideally, namely
$\theta_V=\theta_V^\textrm{ideal}\equiv\arctan\frac{1}{\sqrt{2}}$, the nonet
symmetry breaking strength is $\delta_E\equiv r_E-1 = (+21.0\pm1.6)\%$
or $(-16.4 \pm 1.0)\%$ ($(+3.9 \pm 0.8)\%$ or $(-3.7 \pm 0.7)\%$) for
Solution I (II). On the
other hand, we obtain $\phi_V \equiv
|\theta_V-\theta_V^\textrm{ideal}|=4.97^\circ\pm0.33^\circ$
($1.03^\circ\pm0.19^\circ$) for Solution I (II) assuming nonet
symmetry. However, $\phi_V$ is
found to be $3.84^\circ$ from the quadratic mass formulae~\cite{pdg14}
and $3.34^\circ \pm 0.09^\circ$ from a global fit to the radiative
transitions of light mesons~\cite{KLOE}. The $\phi_V$ values do not
agree with either solution. This is the first indication that nonet
symmetry~\cite{phemodel} is broken and the doubly OZI-suppression process contributes
in $\jp$ electromagnetic decays.

\section{Summary}
Using $\psip$ decays of 106 million, a sample of 1.31 billion $\jp$ events and $\psi(3770)$ data of $2.9~fb^{-1}$ integrated luminosity, many analysis are performed. Exclusively baryonic decays of the $\psipp$ are searched for, but no significant signals are observed, and the upper limits for the branching fractions are set for these decays. For the radiative transition $\psipp\to\gamma\etacp$, the $\psipp$ transition to $\chicj$, isospin violation decay $\chi_{c0,2}\to\pi^0\etac$, the $C-$parity violation decays $\jp\to\gamma\gamma,\gamma\phi$ are searched for, but no significant signals are observed, and upper limits are set for these decays. The decays of $\psipp\to\gamma\chi_{c1}$ and $\jp\to\pi^0\phi$ signals are observed. These measurements provide more information on the charmonium structure, and the isospin and $C$-parity violation in the charmonium decays.
\Acknowledgements
This work is partly supported by the National Natural
Science Foundation of China under Grants No. 11375205.


\end{document}